\documentclass[aps,prd,onecolumn,groupedaddress,showpacs,nofootinbib,amssymb
]{revtex4}
\usepackage[dvips]{graphicx}
\usepackage{amssymb}
\usepackage{amsmath}
\usepackage{graphicx}
\usepackage{amsfonts}
\usepackage{bm}

\begin{document}

\title{Generalized Logarithmic Equation of State in
Classical and Loop Quantum Cosmology Dark Energy-Dark Matter
Coupled Systems}
\author{V.K. Oikonomou,$^{1,2,3}$\,\thanks{v.k.oikonomou1979@gmail.com}}
\affiliation{$^{1)}$ Department of Physics, Aristotle University of Thessaloniki, Thessaloniki 54124, Greece\\
$^{2)}$ Laboratory for Theoretical Cosmology, Tomsk State
University of Control Systems
and Radioelectronics (TUSUR), 634050 Tomsk, Russia\\
$^{3)}$ Tomsk State Pedagogical University, 634061 Tomsk, Russia }
\tolerance=5000

\begin{abstract}
In this paper we shall study the phase space of a coupled dark
energy-dark matter fluids system, in which the dark energy has a
generalized logarithmic corrected equation of state. Particularly,
the equation of state for the dark energy will contain a
logarithmic function of the dark energy density $\rho_d$ and will
also have quadratic and Chaplygin gas-like terms, expressed in
terms of $\rho_d$. We shall use the dynamical system approach in
order to study the cosmological dynamics, and by appropriately
choosing the dynamical system variables, we shall construct an
autonomous dynamical system. The study will be performed in the
context of classical and loop quantum cosmology, and the focus is
on finding stable de Sitter attractors. As we demonstrate, in both
the classical and loop quantum cosmology cases, there exist stable
de Sitter attractors in the phase space, with the loop quantum
cosmology case though having a wider range of the free parameter
values for which the stable de Sitter attractors may occur. It is
emphasized that the use of a generalized dark energy equation of
state makes possible the existence of de Sitter attractors, which
were absent in the case that a simple logarithmic term constitutes
the dark energy equation of state.
\end{abstract}


\maketitle

\section{Introduction}

The discovery of the late-time acceleration of our Universe in the
late 90's \cite{Riess:1998cb}, is to our opinion the most
surprising and mysterious discoveries ever made for our Universe.
This is due to the fact that no one actually expected this
evolutionary process for our Universe, and in fact it is
counterintuitive from many aspects. Of course it is a theoretical
challenge to explain this mysterious late-time acceleration dubbed
dark energy, and in classical Einstein-Hilbert gravity contexts, a
negative pressure fluid is needed to produce this kind of
evolution for the Universe, which in many cases is a phantom fluid
\cite{Caldwell:2003vq}. Apart from this late-time phenomenon, the
nature of dark matter is not yet determined too, but dark matter
is the basic ingredient of the phenomenologically successful model
called $\Lambda$-Cold-Dark-Matter model. Also dark matter can
successfully explain the galactic rotation curves, however there
exist various shortcomings \cite{Tulin:2017ara}, which at a
galactic level can be explained by changing the equation of state
of dark matter (EoS)
\cite{Berezhiani:2015bqa,Berezhiani:2015pia,Hodson:2016rck,Berezhiani:2017tth}.
As already mentioned, no direct proof exists that can explain the
nature of dark matter, although there exist particle physics
proposals which assume that dark matter consists of weakly
interacting particles \cite{Oikonomou:2006mh}.

For the explanation of dark energy, modified gravity offers one of
the most consistent theoretical frameworks, which can successfully
explain the late-time acceleration
\cite{reviews1,reviews2,reviews3,reviews4,reviews5,reviews6} and
in some cases it is possible to describe both the inflationary and
the late-time acceleration era within the same theoretical
framework, see for example \cite{Nojiri:2003ft}. Apart from the
modified gravity description, there exists a research stream in
the literature of modern theoretical cosmology, which assumes that
the dark sector is composed by two interacting fluids, the dark
matter and the dark energy fluids, see for example Refs.
\cite{Gondolo:2002fh,Farrar:2003uw,Cai:2004dk,Bamba:2012cp,Guo:2004xx,Wang:2006qw,Bertolami:2007zm,He:2008tn,Valiviita:2008iv,Jackson:2009mz,Jamil:2009eb,He:2010im,Bolotin:2013jpa,Costa:2013sva,Boehmer:2008av,Li:2010ju,Yang:2017zjs}.
Actually the fluid cosmological description is quite frequently
adopted for the dark energy description
\cite{Barrow:1994nx,Tsagas:1998jm,HipolitoRicaldi:2009je,Gorini:2005nw,Kremer:2003vs,Brevik:2018azs,Carturan:2002si,Buchert:2001sa,Hwang:2001fb,Cruz:2011zza,Oikonomou:2017mlk,Brevik:2017juz,Brevik:2017msy,Nojiri:2005sr,Capozziello:2006dj,Nojiri:2006zh,Elizalde:2009gx,Elizalde:2017dmu,Brevik:2016kuy,Balakin:2012ee,Zimdahl:1998rx},
usually having a non-trivial EoS. The existence of an interaction
between the dark sector fluids is observationally supported by the
fact that the dark energy dominates over the dark matter component
of our Universe after galaxy formation until the late-time era. In
addition, it is known that the dark matter density $\Omega_{DM}$
cannot be calculated without determining the dark energy density
$\Omega_{DE}$ \cite{Kunz:2007rk}.

Recently we examined the phenomenological consequences of a dark
energy fluid coupled with a dark matter fluid, with the dark
energy fluid having a logarithmic-corrected EoS
\cite{Odintsov:2018obx}, see also \cite{Ferreira:2016goc}. The
presence of logarithmic terms is inspired by solid state physics,
in which the pressure of the deformed crystalline solids under the
isotropic stress has a logarithmic dependence
\cite{Intermetallics1,Intermetallics2,Ivanovskii}. As we
demonstrated, it is possible to have stable accelerating
attractors in the phase space of the interacting dark energy-dark
matter system, and specifically quintessential ones. Actually, we
proved explicitly that there exist several stable quintessential
fixed points, but no de Sitter attractors were found for the
corresponding dynamical system. Motivated by the absence of de
Sitter attractors in the logarithmic corrected coupled dark
energy-dark matter system, in this paper we shall use a
generalized logarithmic corrected EoS for the dark energy, and we
shall study the dynamical evolution of the coupled dark
energy-dark matter system. We shall use the autonomous dynamical
system approach, and we shall extensively study the phase space
structure of the cosmological system, emphasizing on the existence
of stable de Sitter fixed points. Also we shall investigate the
stability of these fixed points, by using a numerical approach,
due to the fact that the resulting algebraic equations cannot be
solved analytically. The dynamical system approach for studying
the dynamical evolution of various cosmological systems is quite
popular in the modern cosmology literature
\cite{Odintsov:2018uaw,Odintsov:2018awm,Boehmer:2014vea,Bohmer:2010re,Goheer:2007wu,Leon:2014yua,Guo:2013swa,Leon:2010pu,deSouza:2007zpn,Giacomini:2017yuk,Kofinas:2014aka,Leon:2012mt,Gonzalez:2006cj,Alho:2016gzi,Biswas:2015cva,Muller:2014qja,Mirza:2014nfa,Rippl:1995bg,Ivanov:2011vy,Khurshudyan:2016qox,Boko:2016mwr,Odintsov:2017icc,Granda:2017dlx,Landim:2016gpz,Landim:2015uda,Landim:2016dxh,Bari:2018edl,Chakraborty:2018bxh,Ganiou:2018dta,Shah:2018qkh,Oikonomou:2017ppp,Odintsov:2017tbc,Dutta:2017fjw,Odintsov:2015wwp,Kleidis:2018cdx,Oikonomou:2019muq,Oikonomou:2019boy}
and in this paper we shall appropriately form an autonomous
dynamical system for the cosmological system at hand. We shall use
two theoretical frameworks, namely the classical Einstein-Hilbert
framework and the loop quantum cosmology (LQC) framework
\cite{LQC1,LQC3,LQC4,LQC5,Salo:2016dsr,Xiong:2007cn,Amoros:2014tha,Cai:2014zga,deHaro:2014kxa,Kleidis:2018plu,Kleidis:2017ftt},
and for each case we shall study the trajectories in the phase
space, the existence of de Sitter fixed points and finally we
shall examine the stability of the de Sitter fixed points. To our
surprise, in both cases the effect of a generalized EoS in the
dark energy fluid leads to stable exactly de Sitter fixed points.
In fact, the EoS behaves exactly as in the case of an exact de
Sitter cosmology. The resulting picture is interesting since for
both the classical and the LQC cases, the existence of de Sitter
fixed points occurs, however in the LQC case, there is a wide
range of free parameters for which the occurrence of stable de
Sitter attractors is ensured. We should also note that the
presence of the logarithmic term is crucial since it stabilizes
the fixed points of the dynamical system, both in the classical
and in the LQC cases.

This paper is organized as follows: In sections II and III we
shall study the classical gravity and LQC coupled dark energy-dark
matter system respectively. Particularly, we shall present the
general form of the dark energy EoS and accordingly, by
appropriately choosing in each case the dynamical system
variables, we shall construct an autonomous dynamical system, and
we study in detail the phase space structure of the cosmological
system. We emphasize on the existence of de Sitter fixed points
and their stability, so by using a numerical approach we prove the
existence of stable de Sitter fixed points in the phase space of
the classical and of the LQC cosmology systems.

Before we get to the core of our work, we shall briefly discuss
the geometric background which shall be assumed in throughout in
this paper. Particularly, we shall consider a flat
Friedmann-Robertson-Walker (FRW) spacetime with line element,
\begin{equation}\label{frw}
ds^2 = - dt^2 + a(t)^2 \sum_{i=1,2,3} \left(dx^i\right)^2\, ,
\end{equation}
with $a(t)$ being the scale factor of our Universe. Accordingly,
the corresponding Ricci scalar is equal to,
\begin{equation}\label{ricciscalaranalytic}
R=6\left (\dot{H}+2H^2 \right )\, ,
\end{equation}
where $H=\frac{\dot{a}}{a}$ denotes as usual the Hubble rate of
our Universe. Finally, we shall use a physical units system in
which $\hbar=c=1$.

\section{The Classical Cosmology Case: Generalized Logarithmic EoS Effects on the Phase Space}

\begin{figure}[h]
\centering
\includegraphics[width=20pc]{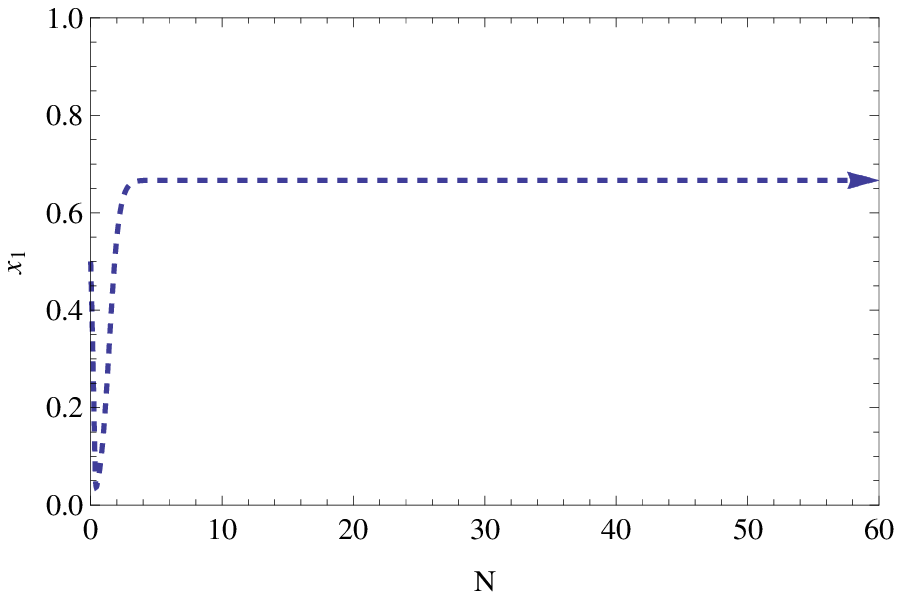}
\includegraphics[width=20pc]{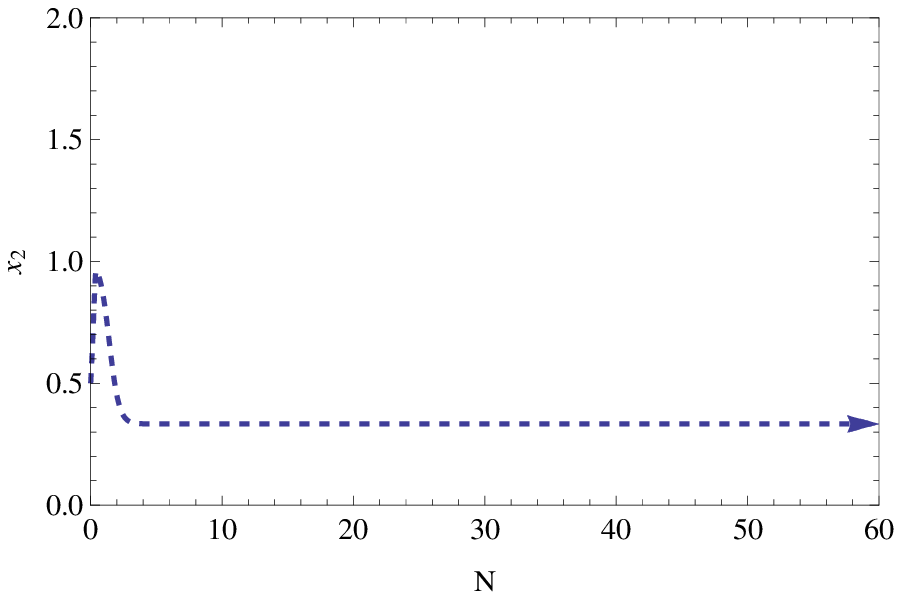}
\includegraphics[width=20pc]{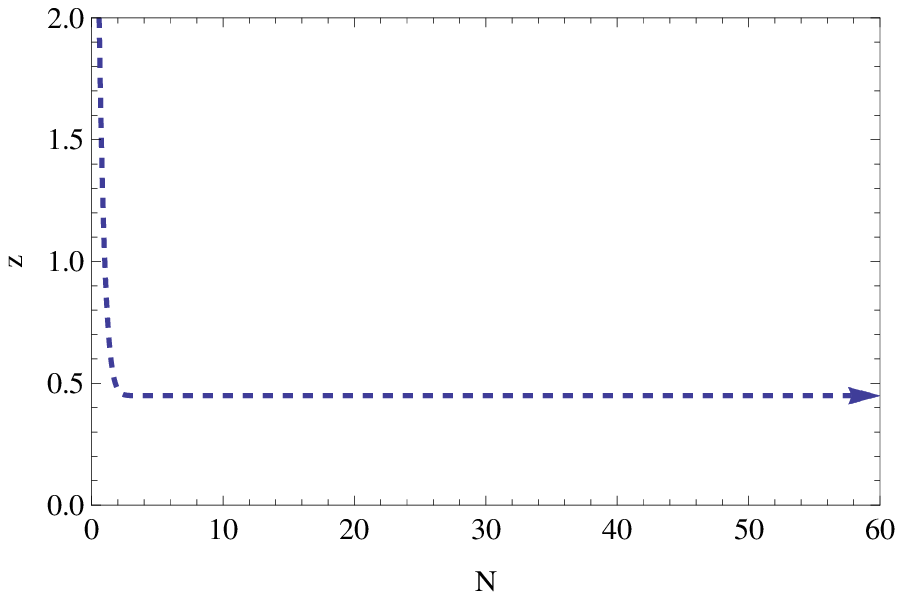}
\caption{{\it{The behavior of the variables $x_1(N)$, $x_2(N)$ and
$z(N)$ as functions of the $e$-foldings number $N$, for the first
60 $e$-foldings. The values of the free parameters are chosen to
be $(\tilde{A},B,c_1,c_2,w_d,\Lambda)=(-1,0,1,1,-\frac{1}{3},1)$.
.}}} \label{plot1}
\end{figure}

\begin{figure}[h]
\centering
\includegraphics[width=20pc]{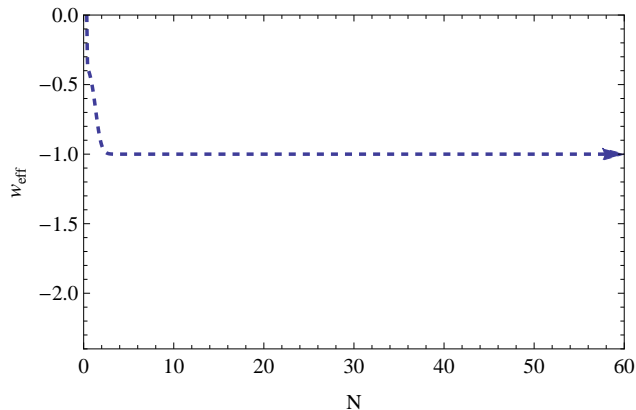}
\caption{{\it{The behavior of the total EoS parameter $w_{eff}$ as
a function of the $e$-foldings number $N$, for the first 60
$e$-foldings, with
$(\tilde{A},B,c_1,c_2,w_d,\Lambda)=(-1,0,1,1,-\frac{1}{3},1)$. As
it can be seen the fixed point is a stable exact de Sitter fixed
point.}}} \label{plot2}
\end{figure}

\begin{figure}[h]
\centering
\includegraphics[width=20pc]{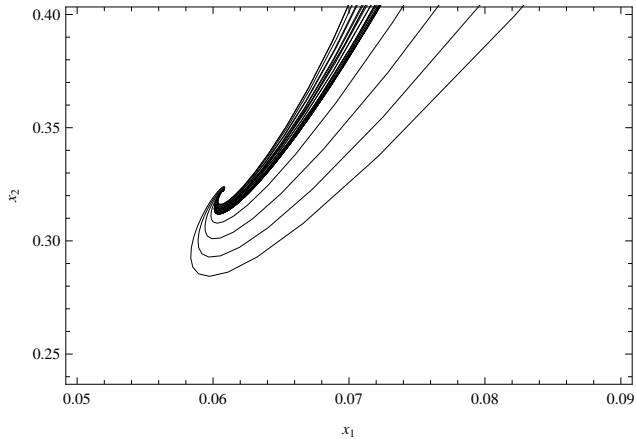}
\caption{{\it{The behavior of the trajectories of the classical
cosmological system phase space in the plane $x_1-x_2$, for
various initial conditions and with
$(\tilde{A},B,c_1,c_2,w_d,\Lambda)=(-1,0,1,1,-\frac{1}{3},1)$. The
existence of a final stable attractor is obvious.}}} \label{plot3}
\end{figure}

The first case we shall consider is the classical Einstein-Hilbert
case of the two interacting dark fluids. Particularly, we shall
introduce a generalized logarithmic EoS for the dark energy fluid
and we shall investigate in detail the effects of the logarithmic
term on the phase space structure of the classical system. The
classical Friedmann equation of the coupled dark energy-dark
matter system in the FRW background is,
\begin{equation}\label{flateinstein}
H^2=\frac{\kappa^2}{3}\rho_{tot}\, ,
\end{equation}
where $\kappa^2=8\pi G$, $G$ is Newton's gravitational constant,
and $\rho_{tot}$ is the total energy density of the coupled
system, which is,
\begin{equation}\label{totealeeos}
\rho_{tot}=\rho_m+\rho_d\, ,
\end{equation}
where $\rho_d$ and $\rho_m$ are the dark energy and dark matter
energy density respectively. Taking into account the conservation
of the energy momentum for the coupled system, and also their
non-trivial interaction, the continuity equations are,
\begin{align}\label{continutiyequations}
& \dot{\rho}_m+3H\rho_m=Q\,
\\ \notag & \dot{\rho}_d+3H(\rho_d+p_d)=-Q\, ,
\end{align}
where $p_d$ stand for the dark energy pressure, and the dark
matter fluid has zero pressure. The interaction term $Q$ controls
the amount of energy transferred in between the dark fluids, and
its sign controls which fluid loses energy, for example in Eq.
(\ref{continutiyequations}), if $Q>0$ the dark energy fluid loses
energy and the dark matter fluid gains energy. For
phenomenological reasons we shall assume that the interaction term
will have the following form,
\cite{CalderaCabral:2008bx,Pavon:2005yx,Quartin:2008px,Sadjadi:2006qp,Zimdahl:2005bk},
\begin{equation}\label{qtermform}
Q=3H(c_1\rho_m+c_2\rho_d)\, ,
\end{equation}
where $c_1$, $c_2$ are real positive constants which are
constrained to have the same sign. By combining Eqs.
(\ref{flateinstein}) and (\ref{continutiyequations}), we obtain,
\begin{equation}\label{derivativeofh}
\dot{H}=-\frac{\kappa^2}{3}\left( \rho_m+\rho_d+p_{tot}\right)\, ,
\end{equation}
where $p_{tot}$ denotes the total pressure of the dark fluids,
which is $p_{tot}=p_d$, due to the fact that the dark matter
pressure is zero. In the previous work \cite{Odintsov:2018obx} we
made the assumption that the dark energy EoS contains a
logarithmic term appropriately chosen, and in this work we shall
make a generalization of the dark energy EoS in order to
investigate the phase space structure of the system, focusing on
the existence of de Sitter attractors, which were absent in the
study \cite{Odintsov:2018obx}. Particularly, we shall assume that
the dark energy fluid has the following EoS,
\begin{equation}\label{darkenergyeos}
p_d=\tilde{A} \kappa ^2 \rho_d \ln \left(\frac{\kappa ^2 \rho_d}{3
H^2}\right)-B \kappa ^4 \rho_d^2-\frac{\Lambda }{\kappa ^8
\rho_d}+(-\rho_d) (w_d+1)\, ,
\end{equation}
where $\tilde{A}$, $B$, $\Lambda$ and $w_d$ are real dimensionless
constants. The generalized EoS contains the logarithmic term and
also terms very frequently used in dark energy contexts.
Specifically the term proportional to $\rho_d^2$ is frequently
used in phenomenological dark energy contexts and leads to
singularities when a single dark energy fluid is used
\cite{Nojiri:2005sr}, and also the term $\frac{\Lambda }{\kappa ^8
\rho_d}$ is well known from Chaplygin gas studies, see for example
\cite{Bamba:2012cp,Bento:2002ps,Bilic:2001cg} for standard
references in the field, and also consult Ref.
\cite{Khurshudyan:2018kfk} for a recent work on tachyonic effects.
One crucial and tedious task to perform is to construct an
autonomous dynamical system by using the equations of motion
(\ref{flateinstein}), (\ref{derivativeofh}) along with the
continuity equations (\ref{continutiyequations}) and with the EoS
(\ref{darkenergyeos}). The only way to construct an autonomous
dynamical system from the cosmological equations is to choose the
dimensionless variables of the dynamical system as follows,
\begin{equation}\label{variablesofdynamicalsystem}
x_1=\frac{\kappa^2\rho_d}{3H^2},\,\,\,x_2=\frac{\kappa^2\rho_m}{3H^2},\,\,\,z=\kappa^2H^2\,
,
\end{equation}
and in addition, by using the functional form of the variables
(\ref{variablesofdynamicalsystem}), we can write the interaction
term (\ref{qtermform}) in the following way,
\begin{equation}\label{additionalterms}
\frac{\kappa^2Q}{3H^3}=3c_1x_2+3c_2x_1\, .
\end{equation}
By using Eqs. (\ref{flateinstein}), (\ref{continutiyequations}),
(\ref{derivativeofh}), (\ref{darkenergyeos}),
(\ref{variablesofdynamicalsystem}) and (\ref{additionalterms}) we
can construct the following autonomous dynamical system,
\begin{align}\label{dynamicalsystemmultifluid}
& \frac{\mathrm{d}x_1}{\mathrm{d}N}=-9 B x_1^3 z+9 B x_1^2 z-(c_1
x_2+c_2 x_1)-3 w_d x_1^2+3
w_d x_1+3 x_1 x_2-\frac{\Lambda }{3 z^2} \\
\notag & +3 \tilde{A} x_1^2 \ln (x_1)-3 \tilde{A} x_1 \ln
(x_1)+\frac{\Lambda }{3 x_1 z^2} \, ,
\\ \notag &
\frac{\mathrm{d}x_2}{\mathrm{d}N}=-9 B x_1^2 x_2 z+(c_1 x_2+c_2
x_1)-3 w_d x_1 x_2+3 x_2^2-3 x_2
\\ \notag & +3 A \kappa ^2 x_1 x_2 \ln (x_1)-\frac{\Lambda
x_2}{3 x_1 z^2}\, ,\\ \notag & \frac{\mathrm{d}z}{\mathrm{d}N}=-3
\tilde{A} x_1 z \ln (x_1)+9 B x_1^2 z^2+3 w_d x_1 z+\frac{\Lambda
}{3 x_1 z}-3 x_2 z \, ,
\end{align}
and in addition, the total EoS parameter
$w_{eff}=\frac{p_{d}}{\rho_d+\rho_m}$ can be written in terms of
the variables (\ref{variablesofdynamicalsystem}) in the following
way,
\begin{equation}\label{equationofstatetotal}
w_{eff}=-\frac{-9 \tilde{A} x_1^2 z^2 \ln (x_1)+27 B x_1^3
z^3+\Lambda +9 (w_d+1) x_1^2 z^2}{9 x_1 z^2 (x_1+x_2)}\, .
\end{equation}
From the functional form of the dynamical system
(\ref{dynamicalsystemmultifluid}), it is easy to understand that
finding the fixed points analytically is a formidable task, so we
will rely solely on a numerical approach. Particularly we shall
solve the dynamical system (\ref{dynamicalsystemmultifluid})
numerically, for various sets of initial conditions and for
various values of the free parameters $\tilde{A}$, $B$, $\Lambda$
and $w_d$. After a thorough investigation, with various initial
conditions and several values for the free parameters, it seems
that there is a pattern of behavior in the phase space of the
cosmological system. Particularly, the initial conditions that the
variables $x_1$, $x_2$ and $z$ must such so that the variables
take positive values at some initial time. Secondly, the only case
that leads to a stable fixed point is when $\tilde{A}<0$, $B=0$,
$c_1>0$, $c_2>0$, $w_d<0$ and $\Lambda<0$. Actually, the term
proportional to $B\rho_d^2$ causes strong instabilities in the
phase space, but for the values of the free parameters chosen as
we indicated, the dynamical system has stable fixed points, after
some value of $e$-foldings number. In Fig. \ref{plot1} we present
the behavior of the variables $x_1$, $x_2$ and $z$ as functions of
the $e$-foldings number for the values of the free parameters
chosen as
$(\tilde{A},B,c_1,c_2,w_d,\Lambda)=(-1,0,1,1,-\frac{1}{3},1)$ and
for the initial conditions $x_1(0)=0.5$, $x_2(0)=0.5$, $z(0)=10$.
The plots correspond to the first 60 $e$-foldings.
From Fig. \ref{plot1} it is obvious that a stable fixed point is
reached for quite small values of the $e$-foldings number, and we
can also demonstrate that indeed this is the case. In Table
\ref{table1} we present the values of the variables $x_1$, $x_2$
and $z$ for various values of the $e$-foldings number.
\begin{table*}[h]
\begin{tabular}{@{}crrrrrrrrrrrrrrrrrrr@{}}
\tableline
 $N=1$: $(x_1,x_2,z)=(0.15628,0.84372,0.996944)$.
\\\tableline
$N=5$: $(x_1,x_2,z)=(0.666645,0.333355,0.449231)$.
\\\tableline
$N=10$: $(x_1,x_2,z)=(0.666667,0.333333,0.449231)$.
\\\tableline
$N=60$: $(x_1,x_2,z)=(0.666667,0.333333,0.449231)$.
\\\tableline
\end{tabular}
\small \caption{\label{table1} Values of the variables $x_1$,
$x_2$ and $z$ for various values of the $e$-foldings number.}
\end{table*}
As it can be seen, the fixed point of the dynamical system is
$\phi^{*}=(x_1,x_2,z)=(0.666667,0.333333,0.449231)$, so let us now
investigate the nature of the fixed point. This can be easily done
by evaluating the total EoS parameter $w_{eff}$ by using Eq.
(\ref{equationofstatetotal}), so in Fig. \ref{plot2}, we present
the plots of the total EoS parameter $w_{eff}$ as a function of
the $e$-foldings number $N$, for $N$ chosen in the ranges
$N=[0,60]$. As it can be seen, the total EoS parameter $w_{eff}$
approaches the value $w_{eff}=-1$ after a few $e$-foldings, so the
stable fixed point
$\phi^{*}=(x_1,x_2,z)=(0.666667,0.333333,0.449231)$ we found
numerically, is an exact de Sitter fixed point.
We can further show the existence of an asymptotic attractor in
the phase space of the dynamical system
(\ref{dynamicalsystemmultifluid}), by presenting several
trajectories in the plane $x_1-x_2$, for various initial
conditions. As it can be seen in Fig. \ref{plot3}, there is an
asymptotic attractor for several trajectories which correspond to
different initial conditions. Also in Fig. \ref{plot3} we can also
see a trajectory threading the fixed point. After a closer
analysis it can be shown that this trajectory drives the system to
infinity, so there are initial conditions which may lead to
singularities, however these belong to negative initial
conditions, so we disregard these trajectories.
In conclusion, we demonstrated that the classical coupled dark
energy-dark matter system with the generalized logarithmic dark
energy EoS (\ref{darkenergyeos}) has stable de Sitter attractors.
It is noteworthy that the logarithmic term contributes
significantly to the stabilization of the final attractor, and
also we need to note that the term $\sim B\rho_d^2$ must be set
equal to zero, since it completely destabilizes the phase space.
In the next section we shall investigate whether the LQC effects
can modify or alter the phase space structure of the classical
phase space. As we will demonstrate, the effects of the LQC
theoretical framework are significant since the parameter space
for which the existence of stable de Sitter attractors is
guaranteed, is enlarged.

Before closing, we shall further discuss the stability issue of
the attractors we found in this section, and also we specify the
significance of the logarithmic and other terms in the generalized
EoS (\ref{darkenergyeos}). As it is found by our numerical
analysis, the appearance of the logarithmic term is crucial for
the stability of the final attractors. This feature was also found
in Ref. \cite{Odintsov:2018obx}, however in Ref.
\cite{Odintsov:2018obx}, only the logarithmic term was included,
and the resulting de Sitter attractors were quintessential, not de
Sitter. In the present case with generalized EoS, the final
attractors are stable, an effect guaranteed by the logarithmic
term, and most importantly, the attractors are de Sitter, which is
the effect of the third and fourth term of the generalized EoS
(\ref{darkenergyeos}). Finally it is crucial to note that the
attractors occur only when $B=0$, so the second term of the
generalized EoS (\ref{darkenergyeos}), destabilizes the final
attractors.

\section{The Loop Quantum Cosmology Framework and Interacting Dark Energy-Dark Matter}

\begin{figure}[h]
\centering
\includegraphics[width=20pc]{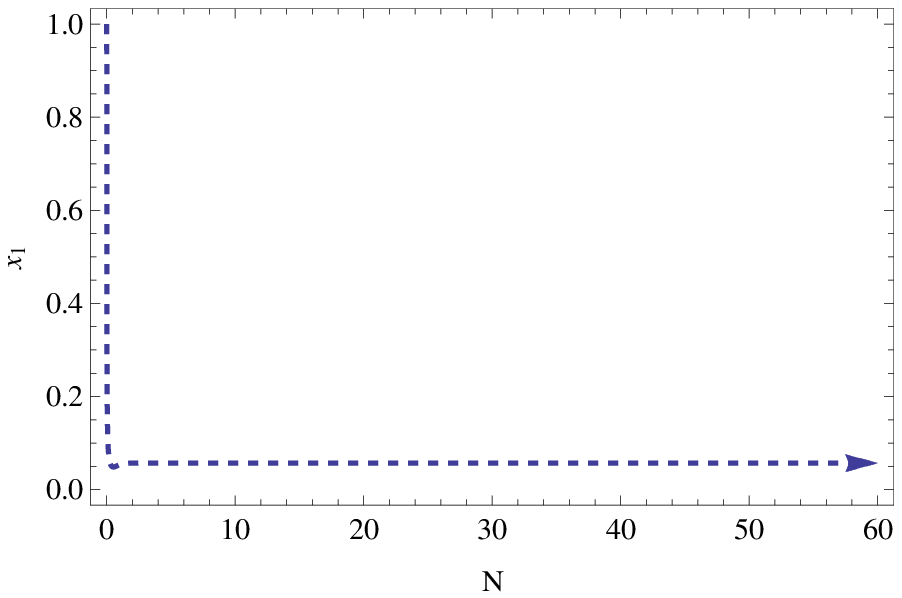}
\includegraphics[width=20pc]{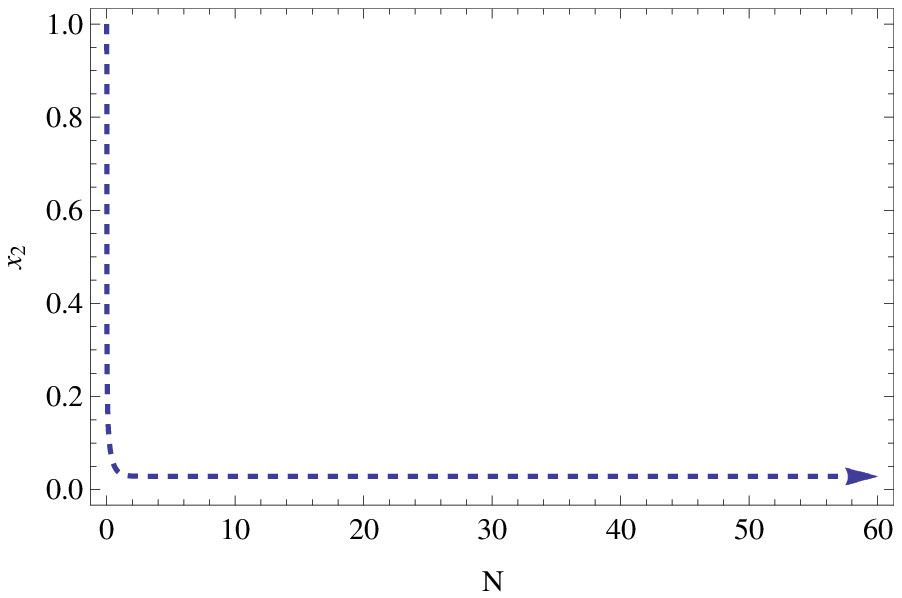}
\includegraphics[width=20pc]{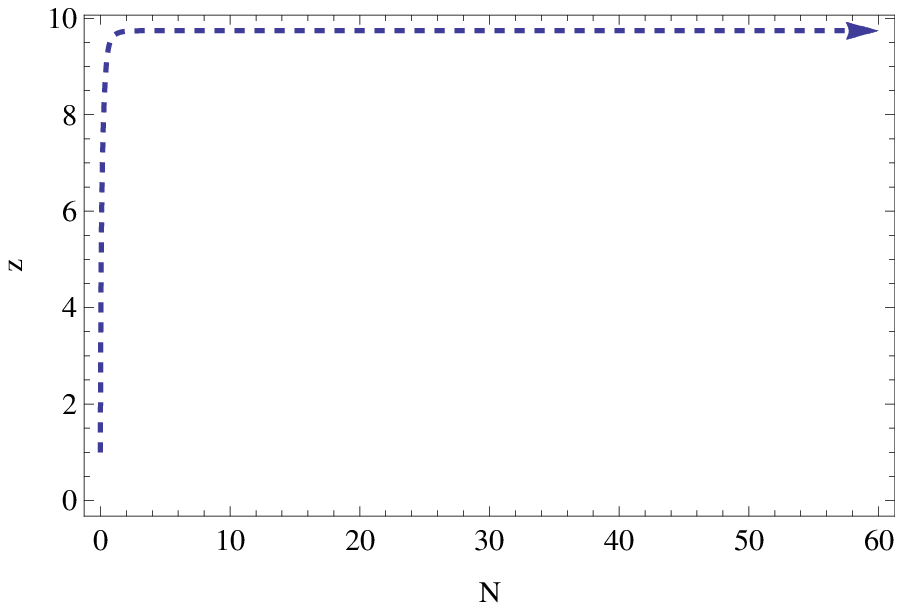}
\caption{{\it{The behavior of the variables $x_1(N)$, $x_2(N)$ and
$z(N)$ as functions of the $e$-foldings number $N$, for the first
60 $e$-foldings. The values of the free parameters are chosen to
be $(A,B,c_1,c_2,w_d,\Lambda)=(1,-1,1,1,-\frac{1}{3},-1)$.}}}
\label{plot4}
\end{figure}

\begin{figure}[h]
\centering
\includegraphics[width=20pc]{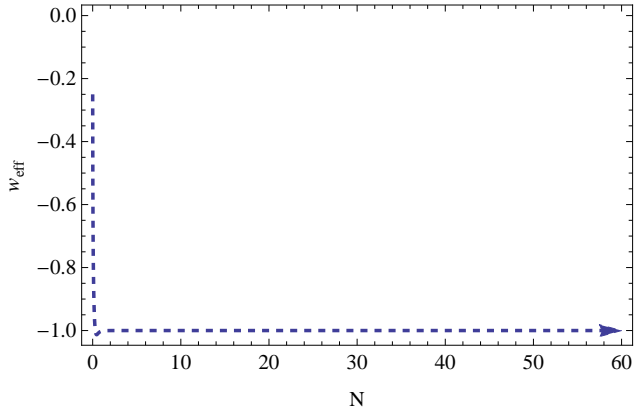}
\caption{{\it{The behavior of the total EoS parameter $w_{eff}$ as
a function of the $e$-foldings number $N$, for the first 60
$e$-foldings, with
$(A,B,c_1,c_2,w_d,\Lambda)=(1,-1,1,1,-\frac{1}{3},-1)$. As it can
be seen the fixed point of the LQC system is a stable exact de
Sitter fixed point.}}} \label{plot5}
\end{figure}

In this section we shall repeat the study we performed in the
previous section, by incorporating LQC effects in the theory. As
we shall see, the LQC effects have a significant contribution to
the resulting picture, since the existence of stable de Sitter
attractors is ensured for a wider range of free variables, and
also we have de Sitter attractors even in the case that $B\neq 0$,
a feature certainly absent in the classical theoretical framework.
The essential features of LQC can be found in various articles,
see for example Refs.
\cite{LQC1,LQC3,LQC4,LQC5,Salo:2016dsr,Xiong:2007cn,Amoros:2014tha,Cai:2014zga,deHaro:2014kxa,Kleidis:2018plu,Kleidis:2017ftt},
so we start of with the LQC Friedmann equation which for the flat
FRW metric of Eq. (\ref{frw}) becomes,
\begin{equation}\label{flateinsteinlqccase}
H^2=\frac{\kappa^2\rho_{tot}}{3}\left(
1-\frac{\rho_{tot}}{\rho_c}\right)\, ,
\end{equation}
where $\rho_{tot}$ is the total energy density
$\rho_{tot}=\rho_d+\rho_m$. The dark energy and dark matter
continuity equations are still given by
(\ref{continutiyequations}) and the interaction term is given by
(\ref{qtermform}). By combining Eqs. (\ref{flateinsteinlqccase})
and (\ref{continutiyequations}), we obtain,
\begin{equation}\label{derivativeofhlqccase}
\dot{H}=-\frac{\kappa^2}{2}\left(\rho_m+\rho_d+p_{tot}\right)\left(
1-2\frac{\rho_m+\rho_d}{\rho_c}\right)\, ,
\end{equation}
with $p_{tot}$ being the total pressure which is again in this
case  equal to $p_{tot}=p_d$. In addition, the dark energy EoS
parameter $w_{eff}$ is assumed to be in this case,
\begin{equation}\label{neweosdarkenergy}
p_d=A \rho_d \ln \left(\frac{\kappa ^2 \rho_d}{3
H^2}\right)-\frac{B \rho_d^2}{\rho_c}-\frac{\Lambda
\rho_c^2}{\rho_d}-\rho_d-\rho_d w_d\, ,
\end{equation}
where $B$, $w_d$ and $A$ are dimensionless parameters. In order to
construct an autonomous dynamical system in the LQC case, we
choose the variables of the dynamical system as follows,
\begin{equation}\label{variablesofdynamicalsystemlqccase}
x_1=\frac{\kappa^2\rho_d}{3H^2},\,\,\,x_2=\frac{\kappa^2\rho_m}{3H^2},\,\,\,z=\frac{H^2}{\kappa^2\rho_c}\,
.
\end{equation}
Accordingly, in terms of the variables
(\ref{variablesofdynamicalsystemlqccase}), the total EoS parameter
$w_{eff}$ is written as follows,
\begin{equation}\label{equationofstatetotallqccase}
w_{eff}=-\frac{-9 A x_1^2 z^2 \ln (x_1)+27 B x_1^3 z^3+\Lambda +9
(w_d+1) x_1^2 z^2}{9 x_1 z^2 (x_1+x_2)}\, .
\end{equation}
Thus by using Eqs. (\ref{flateinsteinlqccase}),
(\ref{derivativeofhlqccase}), (\ref{continutiyequations}), and
(\ref{variablesofdynamicalsystemlqccase}), after some extensive
algebraic manipulations, the autonomous dynamical system in the
LQC case reads,
\begin{align}\label{dynamicalsystemmultifluidlqccase}
& \frac{\mathrm{d}x_1}{\mathrm{d}N}=-18 A x_1^3 z \log (x_1)-18 A
x_1^2 x_2 z \log (x_1)+3 A x_1^2 \log
(x_1)-3 A x_1 \log (x_1)+54 B x_1^4 z^2\\
\notag & +54 B x_1^3 x_2 z^2-9 B x_1^3 z+9 B x_1^2 z-c_1
x_2-c_2 x_1+18 w_d x_1^3 z+18 w_d x_1^2 x_2 z\\
\notag &-3 w_d x_1^2+3 w_d x_1-18 x_1^2 x_2 z-18 x_1 x_2^2 z+3 x_1
x_2+\frac{\Lambda }{3 x_1 z^2} +\frac{2 \Lambda x_1}{z}+\frac{2
\Lambda x_2}{z}-\frac{\Lambda }{3 z^2} \, ,
\\ \notag &
\frac{\mathrm{d}x_2}{\mathrm{d}N}=-18 A x_1^2 x_2 z \log (x_1)-18
A x_1 x_2^2 z \log (x_1)+3 A x_1 x_2 \log (x_1)\\
\notag &+54 B x_1^3 x_2 z^2+54 B x_1^2 x_2^2
z^2-9 B x_1^2 x_2 z+c_1 x_2+c_2 x_1\\
\notag &+18 w_d x_1^2 x_2 z+18 w_d x_1 x_2^2 z-3 w_d x_1
x_2+\frac{2 \Lambda x_2^2}{x_1 z}-18 x_1 x_2^2 z-\frac{\Lambda
x_2}{3 x_1 z^2}-18 x_2^3 z+3 x_2^2+\frac{2 \Lambda x_2}{z}-3 x_2\,
,
\\ \notag & \frac{\mathrm{d}z}{\mathrm{d}N}=18 A x_1^2 z^2 \log (x_1)+18 A x_1 x_2 z^2 \log (x_1)
-3 A x_1 z \log (x_1)-54 B x_1^3 z^3-54 B x_1^2 x_2 z^3+9 B x_1^2
z^2\\ \notag &-2 \Lambda -18 w_d x_1^2 z^2-18 w_d x_1 x_2 z^2 +3
w_d x_1 z-\frac{2 \Lambda  x_2}{x_1}+18 x_1 x_2 z^2+\frac{\Lambda
}{3 x_1 z}+18 x_2^2 z^2-3 x_2 z \, .
\end{align}
The dynamical system (\ref{dynamicalsystemmultifluidlqccase}) is
an autonomous dynamical system, however it is quite complicated to
study it analytically, so we will rely to numerical analysis
again. After a thorough investigation of the free parameters
space, the resulting picture is more rich in comparison to the
classical case, since the existence of a stable de Sitter
attractor occurs for a wider range of the free parameters values.
For example, we found the following class of parameter values
which guaranteed the existence of a stable de Sitter attractor,
\begin{align}\label{desitterattractorcases}
&
A>0,\,\,\,c_1>0,\,\,\,c_2>0,\,\,\,w_d<0,\,\,\,\Lambda<0,\,\,\,B=0\,
, \\ \notag &
A<0,\,\,\,c_1<0,\,\,\,c_2<0,\,\,\,w_d>0,\,\,\,\Lambda>0,\,\,\,B<0\,
, \\ \notag &
A>0,\,\,\,c_1>0,\,\,\,c_2>0,\,\,\,w_d<0,\,\,\,\Lambda<0,\,\,\,B<0\,
, \\ \notag &
A>0,\,\,\,c_1>0,\,\,\,c_2>0,\,\,\,w_d<0,\,\,\,\Lambda=0,\,\,\,B=0\,
,
\end{align}
and there are more combinations not listed here, that yield
similar phenomenological behavior. From Eq.
(\ref{desitterattractorcases}) we can readily spot two major
differences of the LQC case, in comparison with the classical
picture, firstly, the stable de Sitter final attractor occurs for
positive values of the parameter $A$ and secondly, the parameter
$B$ can also take non-zero values. These cases were absent in the
classical approach, since positive values of the parameter $A$
were not allowed, and also non-zero values of the parameter $B$,
utterly destabilized the phase space. Let us study the phase space
structure for one of the above cases, so let us choose for example
the following set of values for the free variables
$(A,B,c_1,c_2,w_d,\Lambda)=(1,-1,1,1,-\frac{1}{3},-1)$, and in
Fig. \ref{plot4} we plot the behavior of the variables $x_1$,
$x_2$ and $z$ for the first 60 $e$-foldings.
From the plots of Fig. \ref{plot4} it is obvious that a stable
fixed point is reached quite fast. We can find numerically which
is the fixed point, so in Table \ref{table2}, we present the
values of the variables $x_1$, $x_2$ and $z$ for various values of
the $e$-foldings number.
\begin{table*}[h]
\begin{tabular}{@{}crrrrrrrrrrrrrrrrrrr@{}}
\tableline
 $N=1$: $(x_1,x_2,z)=(0.0535404,0.0360011,9.63123)$.
\\\tableline
$N=5$: $(x_1,x_2,z)=(0.0571488,0.0285748,9.74263)$.
\\\tableline
$N=10$: $(x_1,x_2,z)=(0.057149,0.0285745,9.74264)$.
\\\tableline
$N=60$: $(x_1,x_2,z)=(0.057149,0.0285745,9.74264)$.
\\\tableline
\end{tabular}
\small \caption{\label{table2} Values of the variables $x_1$,
$x_2$ and $z$ for various values of the $e$-foldings number for
the LQC case.}
\end{table*}
As it can be seen in Table \ref{table2}, the fixed point of the
dynamical system in the LQC case is
$\phi^{*}=(x_1,x_2,z)=(0.057149,0.0285745,9.74264)$. As in the
classical case, the fixed point is a de Sitter fixed point, as it
can be seen in Fig. \ref{plot5}, where we present the functional
dependence of the total EoS parameter $w_{eff}$ as a function of
the $e$-foldings number $N$, for $N$ chosen in the range
$N=[0,60]$. It is obvious from Fig. \ref{plot5} that the total EoS
parameter approaches quite quickly the de Sitter value
$w_{eff}=-1$. In effect, the fixed point
$\phi^{*}=(x_1,x_2,z)=(0.057149,0.0285745,9.74264)$ is an exact
stable de Sitter fixed point.
So the LQC case of the coupled dark energy-dark matter
cosmological system has more phenomenological interest in
comparison to the classical one, due to the fact that, the terms
that caused instabilities in the classical case, are allowed in
the LQC case, and also these can provide a qualitatively
interesting phenomenology. Also the free parameters allowed values
are significantly more in number in comparison to the classical
case, so in the LQC case, a qualitatively more interesting
phenomenology is obtained.

\section{Conclusions}

In this paper we investigated the phase space of a coupled dark
energy-dark matter cosmological system, in which the dark energy
EoS has a generalized functional form containing logarithmic,
quadratic and Chaplygin gas-like terms of the energy density. We
used two theoretical contexts, namely that of classical
Einstein-Hilbert gravity and that of LQC, and we focused the phase
space study on the existence and stability of stable de Sitter
attractors. After appropriately choosing the variables, we
constructed an autonomous dynamical system for both the classical
and the LQC cases. With regard to the classical case, we
demonstrated that there exist values of the free parameters for
which de Sitter attractors exist. By using a numerical approach,
we showed that the fixed points are actually stable de Sitter
attractors. With regard to the LQC case, we also demonstrated that
stable de Sitter attractors exist, and these occur for a wider
range of the free parameters values. Actually, for the classical
case, we showed that when quadratic terms of the form $\sim
\rho_d^2$ exist in the dark energy EoS, the corresponding
dynamical system does not have stable de Sitter attractors, and
actually the phase space trajectories become strongly
destabilized. In the LQC case, this phenomenon does not occur,
since the terms proportional to $\sim \rho_d^2$ are allowed and
can lead to stable de Sitter attractors. Also in the classical
case, the logarithmic term must have a negative sign in order to
have stable de Sitter attractors, but in the LQC case this
constraint is raised and both negative and positive signs of the
logarithmic term can lead to stable de Sitter attractors.

An important issue we did not address is the occurrence of
finite-time singularities in the cosmological system of the
coupled dark energy-dark matter which we studied. Due to the
presence of the logarithmic and Chaplygin gas terms, the
analytical method of the dominant balances used in Refs.
\cite{Odintsov:2018uaw,Odintsov:2018awm} cannot be used in this
case due to the fact that the dynamical system is not polynomial.
Therefore, one should try to address this issue numerically, but
without any analytical results, it is difficult to extract any
useful information from the resulting picture. So approximations
are needed near the finite-time singularities, but this task is
highly non-trivial and difficult to solve for the coupled fluids
system. Perhaps the best strategy is to study the single
logarithmic dark energy fluid case, and examine the behavior near
the singularities. Another issue which we did not address is to
further modify the equation of state and use terms of the form
$\sim \rho_d^n$, with $n$ some positive rational number. This
could have some effect on the classical Einstein-Hilbert system,
so we hope to address this issue in a future work.

\end{document}